\documentclass[usenatbib,usegraphicx]{mn2e}

\newcommand{\kms}{km s$^{-1}$}

\newcommand{\dusty}{$U_{dust}$}
\newcommand{\oiiil}{[O III]$\lambda5007$}
\newcommand{\oiii}{[O III]}
\newcommand{\hb}{H$\beta$}

\title[The Nuclear Outflow in NGC 2110]{The Nuclear Outflow in NGC 2110}

\author[Rosario et al.]
{D.J.~Rosario$^1$, M.~Whittle$^2$, C.H.~Nelson$^3$ and A.S.~Wilson$^4$\thanks{deceased} \\
$^1$Department of Astronomy and Astrophysics, University of California, Santa Cruz, CA 95064; rosario@ucolick.org\\
$^2$Astronomy Department, University of Virginia, Charlottesville, VA 22903; dmw8f@virginia.edu\\
$^3$Physics and Astronomy, Drake University, Des Moines, IA 50311--4505; charles.nelson@drake.edu\\
$^4$Astronomy Department, University of Maryland, College Park, MD 20742}

\begin{document}

\maketitle

\begin{abstract}

We present a HST/STIS spectroscopic and optical/radio imaging study of the Seyfert NGC 2110
aiming to measure the dynamics and understand the nature of the nuclear outflow in the galaxy. 
Previous HST studies have revealed the presence of a linear structure in the Narrow-Line Region (NLR)
aligned with the radio jet. We show that this structure is strongly accelerated, probably by the jet, 
but is unlikely to be entrained in the jet flow. The ionisation properties of this structure are consistent
with photoionisation of dusty, dense gas by the active nucleus. We present a plausible geometrical model
for the NLR, bringing together various components of the nuclear environment of the galaxy.
We highlight the importance of the circum-nuclear disc in determining the appearance of the emission line
gas and the morphology of the jet. From the dynamics of the emission line gas, we place constraints on the accelerating
mechanism of the outflow and discuss the relative importance of radio source synchrotron pressure,
radio jet ram pressure and nuclear radiation pressure in accelerating the gas. 
While all three mechanisms can account for the energetics of the emission line gas, 
gravitational arguments support radio jet ram pressure as the most likely source of the outflow. 

\end{abstract}

\section{Introduction}

Most Active galactic nuclei (AGN) are radio-quiet, i.e, their nuclear radio emission accounts 
for less  than about $5$\% of the bolometric luminosity of the accreting nucleus.
However, many are still associated with non-thermal radio sources 
and, in some cases,  these sources can be extended on scales of hundreds of 
parsecs and show collimated bi-polar jet morphologies reminiscent of the jets
seen in the more powerful class of Radio Galaxies. \citep[e.g.][]{nagar99}. 
Among Seyfert galaxies, those that exhibit extended radio sources frequently
display accelerated emission line kinematics \citep{whittle92}, which is usually interpreted 
as the influence of the fast moving jet outflow on the ionised gas in the
Narrow-Line Region (NLR)  \citep[e.g.,][]{whittle88, capetti96, fws98, cooke00, cecil02, whittle04}.
 
Ionisation studies of Seyferts generally support nuclear UV and X-ray radiation as the
principal cause of NLR ionisation, even in strongly jetted Seyferts \citep[e.g., Mkn 78 - ][]{whittle05}.
Models of NLR clouds with dust suggest that radiation pressure from the accreting nuclear source
plays a major part in altering the pressure equilibrium of such clouds and determining their 
ionisation conditions \citep{groves04b}. In addition, the energy and momentum deposited into the 
NLR from the nuclear radiation field can dominate the dynamics of the region and can easily account for the energy 
and momentum of most outflows. Therefore, a key question in understanding the properties of 
AGN outflows is the relative importance of radiation force compared to the 
force from a radio jet or wind. In this paper, we present a detailed analysis of the kinematics 
and ionisation of a nuclear outflow in the bright local Seyfert NGC 2110 and, using geometrical
and dynamical constraints, explore and compare the accelerative capability of 
nuclear radiation and ram/synchrotron pressure from the jet in this galaxy.

We adopt a systemic velocity of $cz\,=\,2335$ km s$^{-1}$ for NGC 2110
based on stellar absorption line measurements \citep{nelson95}.
With H$_0 = 72$ km s$^{-1}$ Mpc$^{-1}$, this gives a 
distance/physical scale for the galaxy of $30.7$ Mpc/$147$ pc arcsec$^{-1}$.

\section{The NLR of NGC 2110: Context and Previous Studies} \label{context}

\begin{figure}
\label{c6fig1}
\centering 
\includegraphics[width=1.1\columnwidth]{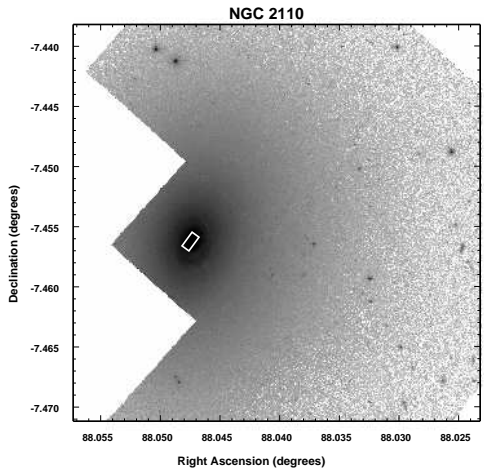}
\caption[Full WFPC2 images of NGC 2110]
{WFPC2 F606W image of NGC 2110 from \citet{malkan98}. The inner 800 pc
region of the galaxy studied in this work and displayed in Fig.~3 is indicated
by a white rectangle}  
\end{figure}

NGC 2110 is a nearby S0 galaxy \citep{devauc91} with a Seyfert 2 AGN \citep{mcclint79}. 
Initially discovered as a strong X-ray 
source \citep{bradt78}, further spectroscopic and radio 
observations \citep{mcclint79,shuder80,ulv83} revealed an 
extended emission line region and radio jet, as well as broadened lines
with FWHM up to $\sim 600$ km s$^{-1}$ in the vicinity of the nucleus,
suggesting a localised jet-ISM interaction.  

Compared to other Seyferts, NGC 2110 has a remarkably well-defined and
symmetric radio jet, with a projected extent of $\sim 2.2''$ ($330$ pc).
The inner jet is approximately linear on both sides of the nucleus 
with a north-south (PA $0^{\circ}$) axis, but bends smoothly 
by about $40^{\circ}$ at a projected distance of $\sim 1''$, 
ending in large bright lobe-like structures. High-resolution VLBA imaging 
shows a parsec-scale feature aligned with the extended inner
jet \citep{mundell00}.

Optical continuum images show a dusty circum-nuclear gas disk
with a well-defined spiral pattern \citep[][ also see panel 5 in 
Fig.~3 of this paper]{gonz02}. Such disks are quite common
in Seyfert galaxies \citep{regan99,pogge02} and their spiral arms
are postulated to be shocks which trigger gas inflow into the active nucleus.
As we discuss later, the presence of a circumnuclear gas disk
may have important consequences for the morphology of the 
jet and the extended ionised gas.

Early ground-based studies of emission line kinematics in NGC 2110
\citep{wilson85a,wilson85b} found evidence for rotation
with the peculiar property that the kinematic centre was offset to the 
south of the continuum peak by $1\farcs7$. 
This apparent paradox has been clarified by more recent observations. 
Ground-based integral-field spectroscopy across the central $10''$
\citep{gonz02,ferruit04}) has shown that the stellar velocity field is rotationally symmetric about the
optical continuum peak (the true nucleus). The gas kinematics, however, is
confirmed to be asymmetric, though the origin of this asymmetry is still unclear.
\citet{ferruit04} argue that the gas in the northern ELR is disturbed by the jet outflow,
suppressing and lowering its average rotational velocity.
 

Our study concentrates on the central $2''$ (300 pc) and uses high
spatial and spectral resolution data to probe the properties of the
inner NLR which, as we show in \S4, is most influenced by the outflow.
In \S7.5, we discuss our results in the context of the gas flows on
kpc-scales.


\section{Datasets: Reductions and Measurements}  \label{data}

\begin{table*}
\caption{HST/STIS Spectroscopic Observations of NGC 2110}

\small

\begin{tabular}{llll}
\hline 

&&Grating&G430M\\
&&$\lambda$ range (\AA) :&4818$-$5104 \\
Aperture&&$\Delta\lambda$ (\AA\ pix$^{-1}$) :&0.277 \\
&&$\Delta$cz (km s$^{-1}$ pix$^{-1}$) :&17 (@ $\lambda5007$) \\
\hline
NGC 2110 A&Ap=$52\arcsec\times 0\farcs2$&Dataset :&O5G401010\\
&PA=-36.202$^{\circ}$&Date :&24/12/2000\\
&Offset=$0\farcs0$, $0\farcs0$&Exposure time :&1522\\
$^{\phantom{1}}$\\
\\
NGC 2110 B&Ap=$52\arcsec\times 0\farcs2$&Dataset :&O5G401020\\
&PA=-36.202$^{\circ}$&Date :&24/12/2000$^{\phantom{1}}$\\
&Offset=$-0\farcs54$, $-0\farcs398$&Exposure time :&600
$^{\phantom{1}}$\\

\hline 

\end{tabular}
\end{table*}

\subsection{STIS spectroscopy}


Two high S/N STIS long-slit spectra were taken at nuclear and off-nuclear positions
in order to cover as much of the central emission line region as possible. 
We used the G430M grating to the [OIII]$\lambda\lambda 4959, 5007$ 
doublet and H$\beta$ at a velocity resolution of FWHM $ \approx 35$ \kms.
See Table 1 for details.
The [OIII]$\lambda 5007$ line tracks the kinematic properties of the emission line gas, 
while the [OIII]/H$\beta$ ratio is sensitive to its ionisation state, 
relatively free of reddening. Henceforth, we will use the labels Slit A and Slit B 
for the nuclear and off-nuclear slit respectively. 

We also use an archival STIS G750M (red) spectrum, first presented in \citet{ferruit04}, which 
covers the lines of [OI]$\lambda 6300$, H$\alpha$, [NII]$\lambda\lambda 
6548,6583$ and [SII]$\lambda\lambda 6717,6731$. Like our nuclear G430M 
spectrum, the slit was centred on the continuum peak of the galaxy, though
along a PA of $155.65^{\circ}$. Fig.~2 compares the nuclear slit 
positions of the G430M and G750M datasets. The angular
difference of $11.8^{\circ}$ between the slits ensures sufficient overlap of the
apertures across the central one arcsecond of the NLR to allow a direct
comparison of line measurements from the red and green spectra
in the inner NLR. 

For all the STIS data, spectrophotometric calibration was 
done using the  CALSTIS pipeline and any remaining hot pixels were 
cleaned using the IRAF `COSMICRAYS' task. 


Emission line properties were measured from one-dimensional 
extractions of the two-dimensional STIS spectrum. The extraction widths were
from one to several spatial increments and were designed to maximise 
the S/N in the emission lines, while isolating regions with coherent kinematic 
behaviour. For each 1-D spectrum, we fitted a smooth polynomial to 
the underlying continuum subtracted it to obtain a pure emission line spectrum. The emission
line strengths were not corrected for any underlying absorption, which is
negligible given the very high equivalent widths of the lines
in the spectra ($> 1000$ across the entire NLR).  
 
We measured line kinematics from the [OIII]$\lambda5007$ line
in the G430M spectra, while emission line ratios were estimated
using the [OIII]$\lambda5007$ line as a scalable template. Our complete
methodology is essentially the same as that applied to a similar
dataset for Mkn 78 \citep{whittle05} and we refer the reader
to this paper for further details.

\subsection{HST and Radio Images}

\begin{figure}
\label{c6fig2}
\centering 
\includegraphics[width=\columnwidth]{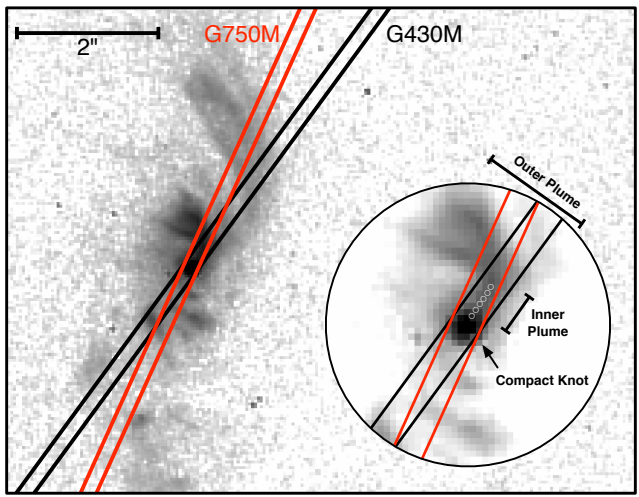}
\caption[The overlap between STIS apertures for the green and red spectral datasets]
{The overlap between the STIS long-slit apertures for the G430M 
and G750M datasets. The apertures are plotted against
the H$\alpha+$[NII] emission line image (see \S3.2 for details). 
North to the top and East to the left. The circular inset is a zoom of the central
2" and certain features identified in \S4 have been indicated. 
Note that the two slit apertures overlap considerably within a nuclear radius of $\sim 0\farcs5$, 
allowing us to adequately combine line measurements from this region (the inner emission line plume)
 in our ionisation study (\S5). The approximate locations of the six bins along each slit that were used
 in this study are shown as small white circles.}
\end{figure}


Two emission line maps of NGC 2110  were prepared from archival HST
imaging datasets. The first is a map of the 
[OIII]$\lambda\lambda4959,5007$+H$\beta$ lines
constructed from WF/PC-1 narrow/medium-band 
images (Program: GO 3724, PI: Wilson). 
The details of the instrument configurations, image reduction 
and analysis are described in \citet{mul94}. 
In brief, following basic calibration with the WFPC
Pipeline, both sets of images were deconvolved with a model 
TINY TIM PSF using a Lucy-Richardson algorithm
to remove the effects of the spherical aberration of HST.
Nearby medium-band continuum images, also appropriately deconvolved,  
were then subtracted to get the final pure emission line images.
The plate scale of the images is $44$ mas/pixel. 

A second emission line map, in H$\alpha+$[NII], was constructed
from archival HST images in the FR680P15 linear ramp filter (`on-band')
and F791W filter (`off-band') taken on the WFPC2 PC1 camera. 
Calibration and reduction details can be found in \citet{ferruit04}. 
This map has significantly better S/N than the WF/PC-1 map and 
reveals faint extended features in the Extended Emission 
Line Region (EELR). In addition, the F791W (approx. I-band) image
provides an emission-line free view of the dust and stellar continuum
geometry around the nucleus.

For brevity, we will refer to the [OIII]$\lambda\lambda4959,5007$+H$\beta$
map simply as the `[O III]' map in the rest of this paper. Similarly, the 
H$\alpha+$[NII]$\lambda\lambda6584,6548$ map will be shortened
to the `H$\alpha$' map. 

    
A VLA X-band radio map of NGC 2110 at 8.4 GHz (3.6 cm), first
presented in \citet{nagar99}, was kindly re-reduced for our purposes 
by N. Nagar. The maps were constructed with uniform weighting and 
self-calibrated with a robustness parameter of 2, 
giving the best compromise between spatial resolution and S/N. 

\subsection{Registration and Astrometry}

Accurate spatial registration of the images, radio maps and STIS slit positions
is essential for this study. The high spatial resolution of HST
warrants absolute positional accuracies on the scale of tens of milliarcseconds. 
This cannot be achieved from the HST astrometric header information alone, since typical
absolute astrometric uncertainties in the HST focal plane are
$\sim1''$. Instead, we employ a number of techniques to register the
various datasets to each other and then anchor them all to the 
astrometric frame of the radio maps, which is accurate to $\sim1$ mas.

Each STIS spectroscopic observing sequence starts with an acquisition image, which, for
our purposes, was taken through the F28X50LP (long-pass) filter. The
relative registration between the slit position and the peak-up centroid
of the acquisition image is better than 5 mas or 0.1 STIS pixels. 
The S/N of these images is sufficient for them to be 
cross-correlated with the WFPC2 F791W images, allowing us to match
the STIS and H$\alpha$ datasets together.

  We assumed  that the nuclear position corresponds to 
the peak of the optical emission in the STIS acquisition images. 
We independently verified this assumption by fitting a gaussian to 
the spatial profile of the stellar continuum in our spectra. 
In all cases, the peak of the continuum profile matched the center 
of the reference pixel along the cross-dispersion axis of the STIS spectra,
as expected for a slit over the peak-up centroid during the target 
acquisition stage of the STIS observations.

    To tie the STIS spectra to the WF/PC-1 images, 
we developed a routine that extracted, for each STIS slit,
a model slit from the [OIII] image and optimised its position and 
PA to match its flux profile to a gaussian-smoothed version of the 
[OIII] flux profile from the STIS spectra.   
The optimisation was performed by minimising the $\chi^{2}$ difference 
between the extracted and real profiles, using a robust Downhill Simplex 
algorithm. With a careful choice of initial parameters to prevent the 
routine from settling into local $\chi^{2}$ minima, we achieved a 
relative registration of  the imaging and spectroscopic datasets to better than 10 mas.

   In order to tie the HST datasets to the radio astrometric frame, 
we registered the centroid of the acquisition image to the co-ordinates 
of the nuclear point source in the corresponding radio map, available
from the literature. This approach is justified by the result from
\citet{ferruit04} that the kinematic center of the stellar velocity field
is the peak in the optical light distribution.
 


\section{Heuristic Description} \label{heurist}

\begin{figure*}
\label{corrplot}
\centering 
\includegraphics[width=0.9\textwidth,angle=0]{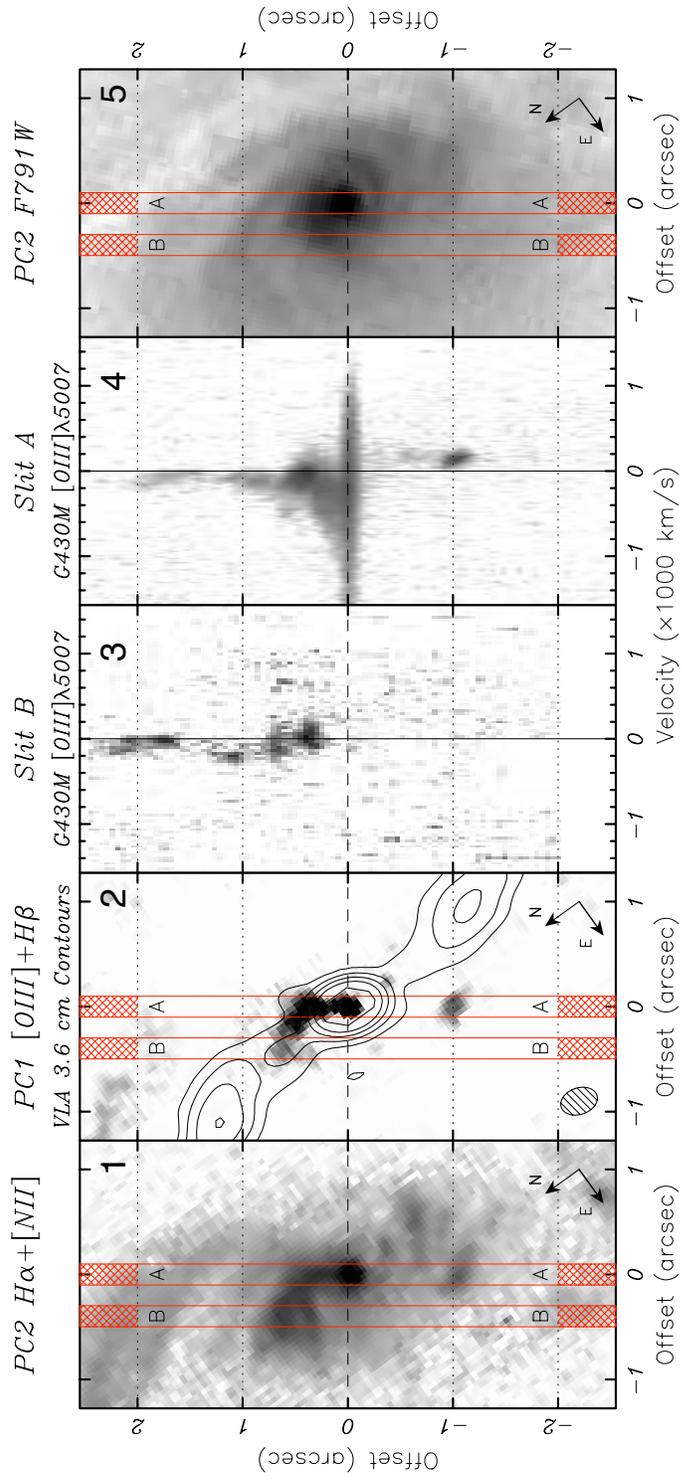}
\caption[NGC 2110: G430M spectra for both slits]
{The [OIII]$\lambda5007$ line for both slits, from the G430M spectra
of NGC 2110. The WFPC2 H$\alpha$, WF/PC-1 [OIII] and WFPC2 F606W images 
are shown for reference. The slit positions are indicated.
Radio contours are over-plotted on the [OIII] image. The hatched ellipse
represents the FHWM beam of the radio map. Dotted lines 
(dashed for the nucleus) are drawn at $1''$ intervals along the
slit direction.}
\end{figure*}

\subsection{Emission Line and Radio Morphology}

The main datasets of NGC 2110 are plotted together in 
Fig.~3. Here, we describe the principal features visible in the data
and discuss the relationships between various identifiable components. 
In what follows, we use the term `NLR' to describe the entire emission
line region, while the term EELR refers only to the outer NLR,
beyond a projected nuclear radius of $\sim 1"$ (150 pc) and out 
to its largest extent of about 4" (600 pc) from the nucleus.

The STIS spectra are sensitive enough  to clearly detect the 
[O III]$\lambda 5007$ line within a nuclear radius of  $2''$. 
Earlier studies have found that the kinematics of the 
[O III]-emitting gas within the inner $1''$ of the NLR are quite disturbed 
and turbulent, while the motion on larger scales is basically dominated
by the gravitational potential of the galaxy \citep{gonz02,ferruit04}.  Therefore, probing these very 
central regions in detail gives us the best insight into the nuclear outflow.

Within $1''$, both the H$\alpha$ morphology (Panel 1 in Fig.~3) 
and [O III] morphology  (Panel 2 in Fig.~3) share some common features. 
The line-emitting gas in both images forms NNW a linear structure 
which curves over in the same direction as the radio jet and the dust lanes 
of the circumnuclear disk (see below). 
The H$\alpha$ image also contains a great deal of extended emission which is not
seen in the [O III] image. This is primarily due to the lower S/N of the [OIII] image,
though some of the difference is a result of the strong extinction towards
the inner NLR south of the nucleus from dust in the circumnuclear disk. 
\citet{ferruit04} show that the high ionisation state of this extended gas 
is indicative of AGN ionisation and not distributed star formation in the disk.

We turn now to the curved structure discussed above. Previous studies have 
referred to this as an emission-line ``jet'', paralleling the structure of 
the radio jet. The STIS spectra of this feature show very strong blue-shifts and complex 
velocity structure, which supports the notion that it is interacting with, or originating
in, the jet. However, the inner emission line ``jet'' has a position angle 
that is offset by $\sim 40^{\circ}$ from the PA of the main radio jet. 
In fact, given that the radio jet preserves its N-S direction down to parsec 
scales \citep{mundell00}, the emission line feature probably skirts the main radio jet, 
rather than being fully cospatial with it. At this juncture, we choose not to assume
that the emission line material is accelerated by the radio jet. As we discuss in \S8,
other mechanisms may also be playing an important role.

To avoid confusion with the radio jet flow, we will henceforth refer 
to this inner emission line structure as a ``plume'' rather than a ``jet''.
For descriptive purposes, we further consider the plume to have two sections: 
an inner and outer plume. The inner plume is the linear high surface brightness 
structure that extends $0\farcs4$ from the nucleus. The outer plume continues from 
the inner plume and curves to the NE. The motivation for separating the 
inner from the outer plume is based on their differing kinematics (\S4.1). 

In order to understand the relationship between the radio and line-emitting
components, it is important to ascertain the role played by dust in modifying
the appearance of the emission line region. At first sight, there is a marked
north-south asymmetry in the emission line distribution with significantly more
emission to the north. However,  Fig.~3, dust lanes are apparent in the
red continuum image just south of the nucleus and suggest extinction
may play a role in creating this asymmetry. 
To constrain the level of extinction, we measure the $H\alpha/H\beta$ Balmer decrement 
south of the nucleus. Despite the misalignment between the G430M and G750M
slits, there is sufficient overlap within $0\farcs5$ of the nucleus
to allow a reliable estimate of Balmer decrement. Across this region, we 
measure $H\alpha/H\beta=6.1\pm4.4$, with large uncertainty due to the faintness of line
emission in this region.  Assuming a Case B value of 3.1, typical of AGN emission line regions, we estimate
a V band extinction of A$_{V}= 1.8^{+1.5}_{-1.8}$ magnitudes, assuming the dust
is distributed as a uniform screen. Using this A$_{V}$ to correct the faint [O III] emission south of 
the nucleus yields a surface brightness comparable to the
northern plume, suggesting that the north and south inner NLR have intrinsically
similar surface brightness. In reality, both the radio and emission line regions
may have a north-south symmetry.

Beyond the bright plume [nuclear radii $>$ 1" (150 pc)], the H$\alpha$ map shows curving arcs
of ionised gas to the north and south, which are, 
for the most part, associated with dusty
spiral arms visible in the continuum images. This EELR is significantly  
extended along PA $\sim 160^{\circ}$, close to that of the disk major axis
(and the line of nodes, assuming a circular disk). The kinematics of this 
extended emission is primarily rotational \citep{ferruit04}. Therefore, this larger-scale ionised 
material is probably disk gas illuminated by the radiation field of 
the central AGN, with dust and higher gas concentrations in the 
spiral arms modulating the appearance of the extended line emission.
While the width of the structure is slightly narrower at 
the nucleus than at its outer extremes, its geometry does not
obviously resemble a canonical double-sided ionisation cone found in many bright 
Seyfert 2s \citep{fws98}. This is not necessarily unusual, since the illumination pattern
of Seyfert nuclei are not always very clear. In \S6, we propose a simple model 
which unifies the geometry of the EELR and the radio jet, and reconciles the illumination pattern 
with a bi-cone. 

\subsection{Emission Line Kinematics}

\begin{figure}
\label{kinematics}
\centering 
\includegraphics[width=\columnwidth]{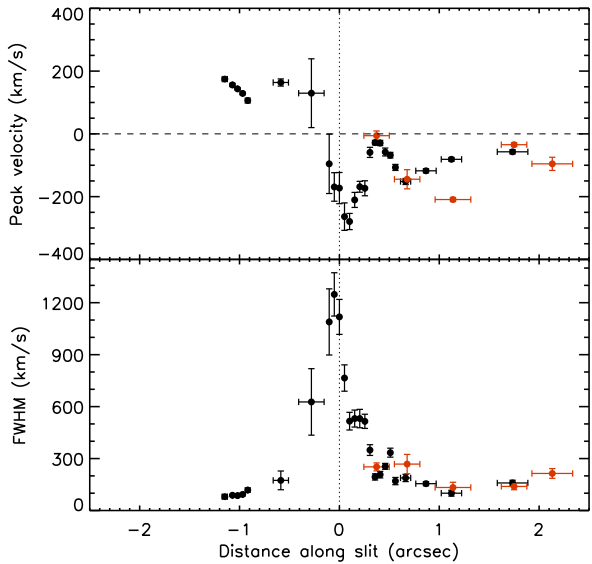}
\caption[NGC 2110: Line Kinematics]
{Peak velocity (top) and FWHM (bottom) of the [OIII]$\lambda 5007$ line as function of position
along the slit, for Slit A (black points) and Slit B (red points).}
\end{figure}

In Fig.~4, we plot peak velocity with respect to systemic (Panel 1)
and FWHM (Panel 2) measured from the [O III]$\lambda5007$ line,  
as a function of nuclear distance along both slits. 
When comparing these measurements to the panels of Fig.~3, 
clear relationships can be seen between the kinematics of the line-emitting gas and its
spatial structure. A compact knot of gas surrounds the nucleus in both [O III]
and  H$\alpha$ maps, resolved with HST, but only marginally resolved
by the STIS spectrograph due to the $0\farcs2$ slitwidth.
This emission line knot displays extremely broad lines 
(FWHM of $1100$ km s$^{-1}$), a median blueshift of 
$250$ km s$^{-1}$ and a predominantly blue wing.  
In addition, in spatial increments south of the nucleus,
a substantial red wing to the [O III] profile becomes evident.
This is consistent with an unresolved bipolar nuclear outflow 
with a dust-extincted red component, smeared out by
the instrumental PSF. 


What might be the origin of the very broad central component?
\citet{moran07} estimate a central supermassive black hole mass in NGC 2110
of M$_{BH} = 2\times10^{8}$ M$_{\odot}$ from its central velocity dispersion
of $\sigma_* = 220$ \kms\ \citep{nelson95}, using the relationship from \citet{tremaine02}.
The radius of gravitational influence for a black hole of this mass given by:

\begin{equation}
R_g \; \approx \; \frac{G M_{BH}}{\sigma_{*}^{2}}
\end{equation}

\noindent which is 18 pc, or 2.4 STIS spatial increments.
While approximate and  dependent on the shape of the potential, 
this estimate of $R_g$ closely matches the size of the nuclear compact knot. 
It is quite possible that some of the high velocities measured in this knot can 
be attributed to gas within a few parsecs of the black hole. On the other hand, there 
appears to be a continuity in the bulk velocity of the gas across the transition 
between the compact knot and the rest of the inner plume,
implying that the plume is the resolved extension of the same
bipolar outflow that extends down into the knot. 
Also, as discussed in \S5.3, the ionisation properties of the knot 
are similar to that of the inner plume. Therefore, the knot probably
contains gas in rotation around the black hole as well as an
outflowing component. We discuss the outflow and its accelerating 
mechanism in \S7.

Beyond the core, the blueshifted gas extends along the entire length of the inner
plume, decelerating with nuclear distance. The  FWHM of the 
gas drops from about $900$ \kms within $0\farcs1$ to systemic 
at the end of the plume where the bulk blue shift
drops to zero. Past the end of the inner plume, the linewidths
rapidly become narrow and the gas returns to a quiescent state, 
consistent with almost normal rotation in the circumnuclear disk.
This velocity pattern suggests that the inner plume is a site of strongly outflowing gas. 
Note that the inner plume lies well outside the sphere of the influence of the black
hole, so its rapid deceleration should not be interpreted as a Keplerian drop-off.


What about the outer plume? Does it inherit the disturbed motion
of the inner plume?  Slit B samples the outer plume as it intersects the central axis of 
the radio jet. This is, in fact, the only part of the plume structure that
appears, in projection, entirely within the confines of the radio jet.
Surprisingly, the ionised material here has narrow linewidths and appears to be 
quiescent and rotational, similar to
the gas beyond the plume in Slit A. In contrast to the inner plume, 
the outer plume is likely to be an AGN-illuminated 
nuclear spiral arm, lying either in front or behind the jet. 
The continuum image (Panel 5  in  Fig.~3) supports this idea: unlike the inner plume,
the outer plume is associated with a similarly shaped
dust feature. As we discuss in \S6, an investigation into the relative geometry
of the NLR and the radio jet provides further evidence that 
the jet is viewed in projection against the outer plume.

Our preliminary analysis of the interaction indicates that the 
inner plume is highly disturbed and directly accelerated
by the nuclear outflow. In the next section, we consider the physical nature 
and excitation of the ionised gas in the nuclear knot and the inner plume.
 
\section{Ionisation Conditions} \label{ionisation}

\begin{figure*}
\label{ratioplot1}
\centering 
\includegraphics[width=\textwidth,angle=270]{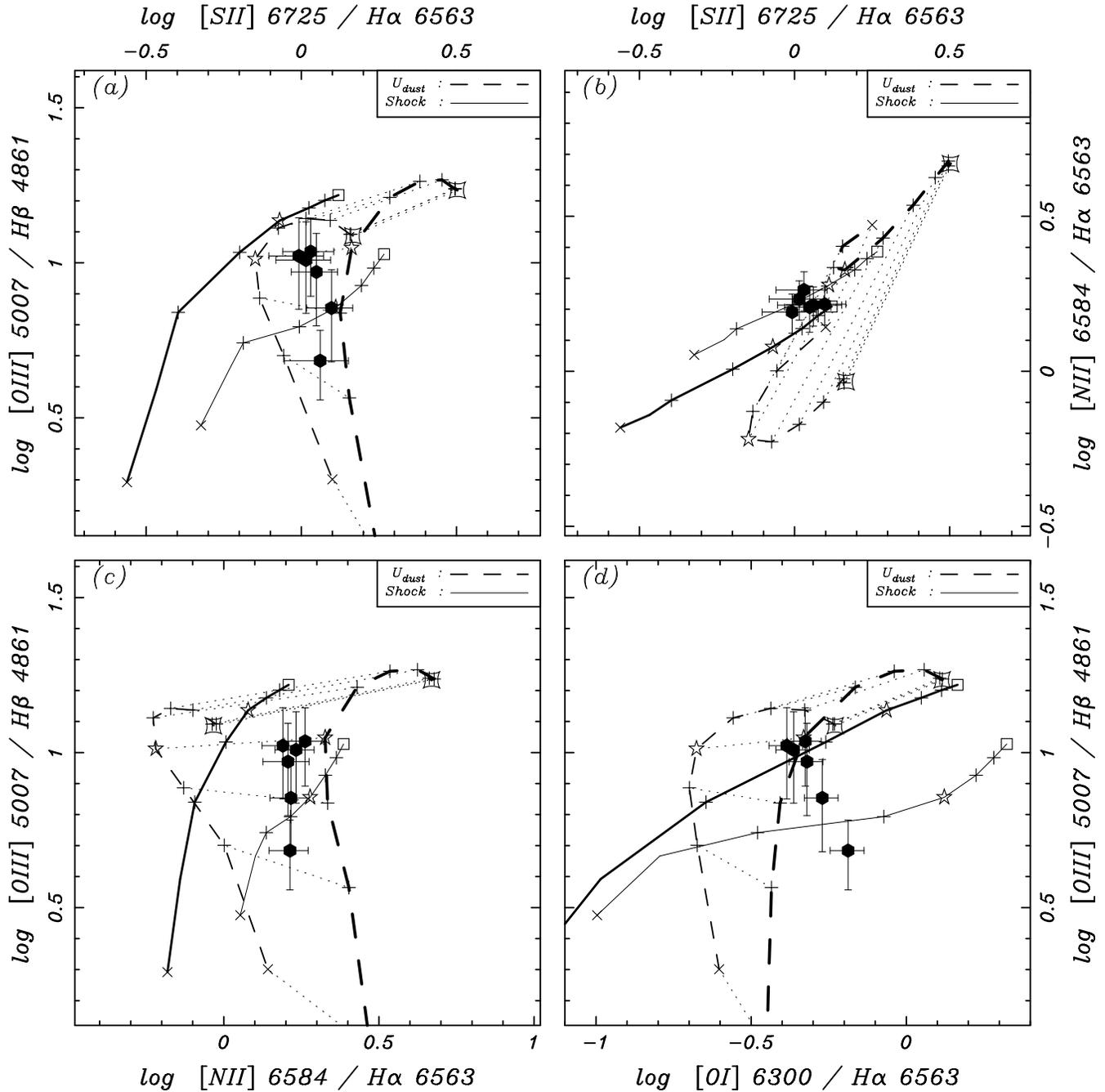}
\caption[NGC 2110: Line ratio-ratio diagrams]
{Emission line ratio-ratio plots for the inner plume in NGC 2110. 
Data points are plotted as filled circles with $2\sigma$ error bars. 
Two sets of ionisation model tracks are plotted: U$_{dust}$ models 
(dashed lines -- sequences in ionisation parameter) and 
shock models from \citet{allen08} (solid lines -- sequences in shock velocity). 
Symbols plotted on each sequence help identify the parameter value.
Shock velocities range from $V_{sh} = 400$ \kms\ (cross) to $1000$ \kms\ (square),
with plus points every 100 \kms\ and a star at 700 \kms. 
$\log U_{dust}$ ranges from $-4.0$ (cross) to $0.0$ (warped square), with a plus point every 0.5 dex
and a star at $-2.5$. Thin/thick lines are pure shock/shock+precursor models for
the shock sequences and $2\times$/$4\times$ solar for the $U_{dust}$ sequences.
More details can be found in  \S5.1.}
\end{figure*}

\subsection{Ionisation models} 

 A knowledge of the ionisation conditions of the emission line gas helps to set the 
stage for a study of the energetics of the inner NLR. The radiation field of the central AGN 
is often a major source of thermal and kinetic energy, through photoionisation 
heating and radiation pressure. In addition, strong shocks, driven into the
NLR clouds by interactions with the radio jet or outflow, can also be a significant source of 
ionising photons \citep{vie88, ds96, allen08} and, under certain conditions, 
this can dominate over nuclear photoionisation. Since shocks are sinks of the
jet kinetic energy flux, constraints from shock ionisation can be used to
determine jet energies. In addition, variations in ionisation state can arise from changes in 
gas density and Lyman continuum optical depth. A comparison
of line strengths with the predictions of models can throw light on the micro-scale nature 
of NLR gas, which then feeds into a dynamical analysis and estimates 
of jet properties.  

We employ two grids of ionisation models applicable to AGN emission line regions. The first 
considers the photoionisation of dusty, optically-thick gas by the EUV and X-ray
continuum of the nuclear source. These models were presented in
\citet{groves04b} and span a range in NLR metallicity, mean ionisation parameter
$U$, gas density and ionising continuum slope. We call these the \dusty\
models. The second set of models are those of autoionising radiative shocks, i.e, shocks
that ionise an optically thin precursor region of unshocked gas, as presented and discussed in  
\citet{allen08}. These models extend the work of \citet{ds96}, spanning a greater
range of shock velocity, gas density, gas metallicity and magnetic compression parameter.
Our approach is to compare measured line ratios with the 
predictions of both of these models, in terms of their
location on line ratio-ratio diagrams and, since we have spatially resolved
the inner NLR,  trends in the measurements. A comparison of 
several different diagrams can allow an assessment of the importance of these
two very  different ionisation mechanisms. 

\begin{table*}

\caption{Emission Line Flux Measurements of the the Inner Plume}
\begin{tabular}{ccccccc}
\hline
Nuc. Dist. & \multicolumn{6}{c}{All fluxes are in units of $10^{-13}$ erg s$^{-1}$ cm$^{-2}$ arcsec$^{-2}$} \\
\cline{2-7}
(arcsec) & \phantom{\LARGE A}H$\beta$\phantom{\LARGE A} & [O III]$\lambda 5007$ & [O I]$\lambda 6300$ & H$\alpha$ & [N II]$\lambda 6584$ & [S II]$\lambda6717+6731$ \\
\hline
0.08  &$0.79 \pm 0.24$    &$3.82 \pm 0.23$    &$2.27 \pm 0.05$      &$3.50 \pm 0.52$    &$5.71 \pm 0.57$      &$4.02 \pm 1.01$  \\  
0.13  &$0.52 \pm 0.21$    &$3.73 \pm 0.22$    &$1.14 \pm 0.05$      &$2.13 \pm 0.32$    &$3.50 \pm 0.35$      &$2.66 \pm 0.40$  \\  
0.18  &$0.40 \pm 0.16$    &$3.76 \pm 0.23$    &$0.79 \pm 0.03$      &$1.66 \pm 0.25$    &$2.67 \pm 0.40$      &$1.85 \pm 0.28$  \\  
0.23  &$0.34 \pm 0.14$    &$3.61 \pm 0.18$    &$0.67 \pm 0.03$      &$1.61 \pm 0.24$    &$2.50 \pm 0.25$      &$1.58 \pm 0.32$  \\  
0.28  &$0.38 \pm 0.15$    &$3.87 \pm 0.19$    &$0.61 \pm 0.04$      &$1.41 \pm 0.21$    &$2.42 \pm 0.24$      &$1.46 \pm 0.29$  \\  
0.33  &$0.38 \pm 0.13$    &$4.10 \pm 0.12$    &$0.67 \pm 0.05$      &$1.43 \pm 0.21$    &$2.61 \pm 0.26$      &$1.53 \pm 0.28$  \\ 

\hline
\end{tabular}

\end{table*}  

\subsection{Line ratio-ratio diagrams}

Fig.~5 plots combinations of optical emission line ratios 
measured for the inner plume. It is worth recalling that these
diagrams compare line ratios from two different STIS datasets,
though any given ratio is constructed from lines measured in a single
dataset (G430M or G750M). The misaligned slits (angular offset of about $12^{\circ}$) 
sample slightly different parts of the inner plume,
from almost full overlap at the nucleus, to approximately 25\% overlap
at the edge of the plume (Fig.~2). Our approach will be valid as long
as there are no strong ionisation gradient perpendicular to the plume,
which seems very unlikely. 

Line ratios, calculated from the flux measurements in Table 2,
are plotted in Fig.~5 as filled circles 
with $2\sigma$ error bars. In what follows, we exclude the set 
of STIS spatial increments on or immediately adjacent to the nucleus to avoid 
complications from the smearing of the nuclear point source 
due to the $0\farcs2$ width of the slit. The approximate central 
locations of these increments are plotted in Fig.~2. As a guide to the reader, the data 
points with lowest [OIII]$\lambda 5007$/H$\beta$ are the closest 
to the nucleus. 

We have compared the full space of parameters from both \dusty\
and shock models against many combinations of line ratios. Fig.~5 is
a distillation of a set of 4 line ratio diagrams and a small subset of model tracks 
that bring out our main results. Plotted using dashed lines, the representative
\dusty\ models feature a power-law ionising spectrum in frequency, 
with a spectral index of $-1.2$ (typical of Seyferts), illuminating
gas clouds with an electron density of $1000$ cm$^{-3}$, with tracks in 
ionisation parameter spanning $-4.0 < \log U < 0.0$ and metallicities of twice and four times solar.
Representative pure shock and shock+precursor models are plotted as solid lines, 
with pre-shock densities of 1 cm$^{-3}$ and shock velocities between 400 and 1000 \kms\, as well
as high magnetic fields 10 mG (highly magnetised shocks). 
We find that in order to match the strengths of the [S II] and [N II] lines, we require 
super solar abundances in both model sets.  
These models were chosen to span the measured line ratios, while
fixing parameters that have a minor influence on the position of the model
tracks on these diagrams. We refer the reader to the original papers 
for a full description of the parameter space covered by the models.


A examination of the line ratio-ratio diagrams yields the
following conclusions:

a.) A gradient in the ionisation sensitive line ratio [O III]/H$\beta$
is observed, towards lower ionisation at smaller radii.
The low ionisation ratios of [N II] and [S II] with H$\alpha$
show no trend and indeed, the entire plume
occupies a very small region of the diagram that 
involves combinations of these three lines (Fig.~5b).
The [O I]/H$\alpha$ line ratio exhibits an anti-correlation with 
[O III]/H$\beta$, which, interestingly, is not mirrored by any
of the model tracks. 

b.) Considering the shock models, the low ionisation points
prefer pure shocks with shock velocities around 700 \kms\ while the high ionisation
points match shock+precursor models with higher velocities. The tracks for unmagnetised
shocks do not span the range of line ratios as well -
weakly magnetic pure shocks produce too weak [O III]/H$\beta$, while
weakly magnetic shock+precursor models produce low values of
[S II], [N II] and [O I] relative to H$\alpha$. Note that the transition from having a significant precursor
contribution near the nucleus to pure shocks at larger radii is required to match the trends 
in the measurements. Neither a full set of pure shock or shock+precursor models actually spans the 
gap between low and high ionisation points.
Models with higher abundances, while not explicitly modeled by \citet{allen08},
are unlikely to improve the match to the measurements, as they will move the [O III] strengths
higher in the shock+precursor models and away from the measurements.

c) On the other hand, the dusty photoionisation models can consistently
reproduce the measured ratios with $\log U \sim 2.0$ and 
abundances about $3\times$solar. There may be a preference for
a harder ionising power-law than we have adopted, which will boost
[S II] and [N II] relative to H$\alpha$, yielding a better fit in Fig.~5b.
A harder ionising continuum will also lead to a better match to the [O I] strengths
in Fig.~5d. However, \citet{groves04b} do not consider 
power-law indices smaller than $-1.2$ and modelling this scenario is beyond
the scope of this paper. In the context of the photoionisation, the
gradient in [O III]/H$\beta$ implies lower ionisation parameters at smaller
nuclear distances. Since the ionising field is expected to decrease with
distance, due to geometric dilution and increasing net optical depth, 
the gradient implies that the density of the ionised gas increases sharply
towards the nucleus.

For a galaxy with the luminosity of NGC 2110 (M$_B = -19.3$), local metallicity scaling relationships 
\citep[e.g.,][]{salzer05} predict oxygen abundances around solar. The inner
kpc of NGC 2110 appears to have significantly higher metalllicities
than the mean galaxy population, though an excess of 0.5 dex 
is not much larger than 1$\sigma$ scatter in abundance for galaxies of this luminosity.

d) Unlike the other ratios, [O I]/H$\alpha$ is not fully consistent with the
best parameters for the \dusty\ sequences (Fig.~5d). Shocks
give a somewhat better match, though, as before, the gradient supports
a switch from pure shocks to shock+precursor emission over the length
of the plume. However, [O I] is a very difficult line to model because
its relative strength is influenced considerably by the high energy (soft X-ray) contribution
to the ionising spectrum, which produce the partially ionised zones
needed to excite neutral oxygen significantly. NGC 2110 has
an extended soft X-ray excess compared to normal Seyferts \citep{weaver95,evans06}
and this may explain why [O I] is stronger than predicted by the photoionisation models.

\subsection{Global Constraints}

A simple check on the feasibility of nuclear photoionisation can be
made by comparing the mid/far IR luminosity of the galaxy with 
the ionising luminosity necessary to power the observed
line emission. Seyferts are known to be powerful IR sources 
with particularly warm colours \citep{rieke78,p_garcia01}. 
Most of the IR emission is believed to come from the 
AGN's optical--UV--X-ray output that is intercepted 
and reprocessed by warm dust near the nucleus and in the torus. 
Since NGC 2110 does not show any signs of strong star formation,
either circumnuclear or in the large-scale galaxy disk, its IR
luminosity $L_{I\!R}$ ($8-1000\mu$m) should be dominated by the output of the AGN.
Using the calibrations of \citet{sanders96}:

\begin{eqnarray}
L_{I\!R}& &\!\! \approx\  1.8\times10^{-11}\, 4 \pi d^{2} \nonumber\\
\times& & (13.5S_{12} + 5.2S_{25} + 2.58S_{60} + S_{100})\ \textrm{ erg s}^{-1}\ \ 
\end{eqnarray}
\vskip 0.2cm
{\noindent
where $S_{12}= 0.35$, $S_{25}=0.84$, $S_{60}=4.13$ and
$S_{100}=5.68$, all in Jy, are the fluxes of NGC 2110 
in the four IRAS bands and $d$ cm is the distance to the galaxy.
This gives us $\log L_{I\!R}\sim 43.7$ erg s$^{-1}$. Assuming that
most of the AGN's radiation comes out at UV to X-ray wavelengths and
is reprocessed, its bolometric luminosity $L_{ph} \approx L_{I\!R}$.
This estimate is consistent with results from spectral fitting of the 
nuclear X-ray emission, appropriately scaled using a bolometric
correction \citep{evans06, moran07}.}

One can estimate the total ionising luminosity using the
total H$\beta$ flux of the Seyfert. Assuming a power-law AGN spectrum of 
the form $L_{\nu} \propto \nu^{-\alpha_o}$ ($\alpha_o > 1.0$),
a dust absorption cut-off at $\lambda_{low}$ and a covering factor 
$\Omega\,C_{i}$ for the ionised gas:

\begin{eqnarray}
L_{H\beta}\ &\approx\ & \Omega\;C_i\, L_{ph} \frac{\alpha_{H\beta}} {\alpha_B} 
\frac{\epsilon_{H\beta}} {\epsilon_H} \frac{\alpha_o -1} {\alpha_o} 
\left(\frac{912} {\lambda_{low}}\right)^{\alpha_o -1} \nonumber\\
&\approx\ & 0.023\,\Omega\, C_i\, L_{ph} \frac{\alpha_o -1} {\alpha_o} 
\left(\frac{912} {\lambda_{low}}\right)^{\alpha_o -1}{\rm erg\ s}^{-1}
\end{eqnarray}
\vskip 0.2cm
{\noindent
where $\alpha_{H\beta}$ and $\alpha_B$ are the H$\beta$ and total case
B recombination coefficients for hydrogen at $T=10^4$K, and 
$\epsilon_{H\beta} = h\nu_{H\beta}$ is the average energy of an 
H$\beta$ photon.}

$\Omega\,C_i$ is the fraction of $4\pi$ steradians intercepted
by line emitting gas. In \S\ref{dyns}, we calculate
the intrinsic cloud covering factor $C_i \sim 1$.
Taking the ionising cone half-angle to be $45^{\circ}$ 
(based on estimates in $\S6$), we calculate $\Omega\,C_i \sim 0.3$.
With canonical values of $\alpha_o = 1.2$, 
$\lambda_{low} = 4000\textrm{\AA}$ and using our estimates of 
$L_{ph}$ and $\Omega\,C_i$, $\log L_{H\beta} \sim 40.8$ erg s$^{-1}$.
Despite the uncertainties in this approach, this estimate compares very well 
with the total extinction-corrected H$\beta$ luminosity of 
$\log L_{H\beta} \sim 40.8$ erg s$^{-1}$
from \citet{wilson85b} based on early ground-based
long-slit work.  

The relative importance of shock ionisation can be tested in a similar fashion.
The Extended Emission-Line Region (EELR: nuclear radii $> 2''$)
is undoubtedly ionised by the central AGN, since the undisturbed gas kinematics
on these scales cannot sustain the fast shock velocities ($>$ 500 \kms) needed to
produce its high level of excitation.  

We concentrate instead on the disturbed inner plume. We can compare
the ionising flux experienced by gas in the plume both from shocks and
the central source. We can estimate the amount of
ionising UV radiation generated per unit shock area, using the following
relation from \citet{ds96}:

\begin{equation}
f_{UV,shock} = 1.11\times10^{-3}\, n\: V_{sh}^{3.04}  \; \; \textrm{erg cm$^{-2}$ s$^{-1}$}
\end{equation}

{\noindent where $V_{sh}$ is the shock velocity in units of $100$ \kms\ and $n$ is the preshock density
in cm$^{-3}$. For $V_{sh} = 500$ \kms\ and $n=1$, $f_{UV,shock} = 0.14$ erg cm$^{-2}$ s$^{-1}$.}

Taking the photon luminosity of the AGN $L_{ph}$ and a 
characteristic distance of  the inner plume ($0\farcs5 \approx 75$ pc), we can estimate
an ionising photon flux per unit area from the nuclear source at the plume 
of $f_{UV,nuc} \approx 75$ erg cm$^{-2}$ s$^{-1}$, almost 3 orders of magnitude
higher than the UV flux from shocks. This suggest that the ionisation of the inner plume is dominated
by the photon output of the nucleus. Taken together with the dominance of nuclear
photoionisation in the EELR, we conclude that shock ionisation contributes very little  
to the line emission in NGC 2110.


\subsection{Line Profiles}

A different handle on the nature of the ionisation comes from
a comparison of line profiles in the emission line plume. 
In Fig.~6 we plot the normalised profiles
of \oiiil\ and \hb\ integrated over the inner plume, the nuclear knot and the
region of weak emission south-east of the nucleus along Slit A.
A measure of the ionisation level as a function of velocity is revealed by 
the relative strengths of \oiii\ to \hb\ across the profiles. 

In all three regions, the \oiii\ line is wider than \hb. In the nuclear knot 
and inner plume, this difference in linewidth takes the form of a
stronger, more highly ionised blue wing on \oiii. 
This blue wing in the nuclear knot forms the base of the 
decelerating inner plume emission and displays a similar 
line ratio (\oiii/\hb $\sim 4.0$), implying a continuity between these regions. 
The line core, with absolute velocities $< 200$ km s$^{-1}$
around systemic, exhibits a low LINER-like ratio (\oiii/\hb $\sim 2.8$). 
In contrast to the inner plume, the \emph{red} half of the \oiii\ line 
in the south-east region may be stronger than \hb\, 
though the low S/N in this region precludes a definite measurement. 
This suggests that a highly extincted plume-like structure also exists in 
the inner south-east NLR, consistent with the results of \S4.1.  

Based on these trends, the following picture presents itself: 
the low velocity line core is emitted by ionised gas that is relatively 
undisturbed and might lie in the plane of the circumnuclear disk. The outflow, in contrast,
is highly ionised and possibly mass loaded from material accelerated
off the low ionisation clouds in the disk. This enhancement of ionisation
could be due to a greater fraction of low density, optically thin gas
in the outflow. As it expands away from the nucleus,  the outflow decelerates rapidly 
and finally merges with the rest of the disk, losing its distinct identity and kinematics. 

To conclude, the line-ratio analysis and the agreement 
between the line and reprocessed ionising luminosities of the AGN 
lends some support for widespread nuclear photoionisation. In addition, 
the presence of strong trends in ionisation across line profiles in the innermost 
plume allow us to differentiate between an undisturbed, LINER-like disk component
photoionised by the AGN, and a more highly ionised outflow component.

\section{Description of the Interaction} \label{desc}

\begin{figure}
\label{profilecomps}
\centering 
\includegraphics[width=0.8\columnwidth]{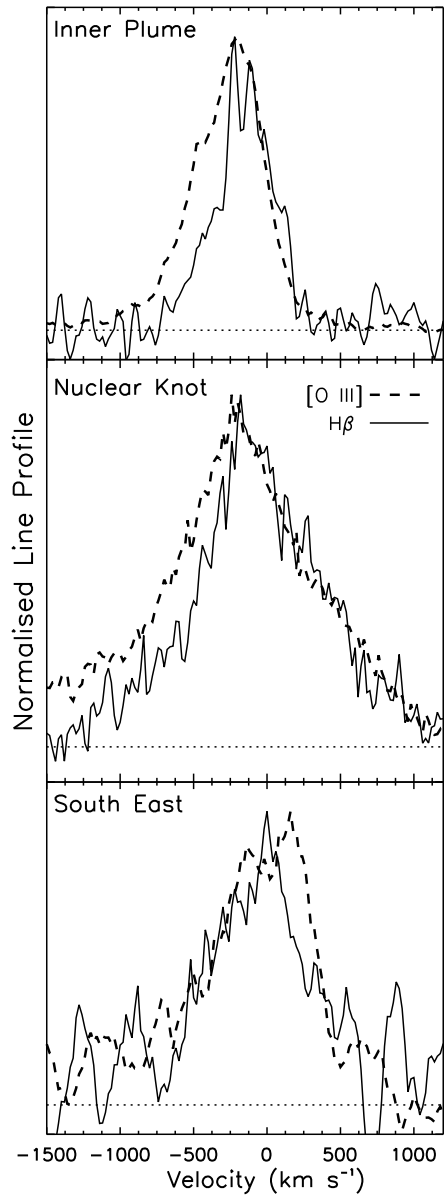}
\caption[NGC 2110: Profile Comparisons]
{A comparison of the [O III]$\lambda 5007$ and H$\beta$ line profiles
in three regions of the inner NLR: the inner plume (top), the nuclear knot (centre)
and the dust obscured region just south-east of the nuclear knot along Slit A (bottom). 
Note the distinctive trend in the upper two panels for the blue wings of the profile
to be stronger in [O III] compared to H$\beta$, implying a high degree of ionisation.
A hint of the reverse (a higher ionisation red wing) is visible in the south-east
inner NLR, but the S/N of the lines is very low here, due to strong dust extinction. 
}
\end{figure}

A geometric and kinematic model for the NLR is necessary for a dynamical
analysis of the interaction. To this end, we first introduce
various structural components of the NLR and then discuss how they relate
to each other.

Based on elliptical isophote fits to the F791W continuum image, 
\citet{gonz02} estimate a circumnuclear disk inclination of 
approximately $42^{\circ}$, with the line of nodes 
aligned along PA $163\fdg5$. The contrast of the
dust lanes indicates that the western side of the disk is 
inclined towards the line of sight. Combining this information 
with the direction of rotation of the ionised gas in our spectra 
(blueshifted to the north and redshifted to
the south), we deduce that the spiral arms \emph{trail} the disk rotation
in the normal fashion of large scale galaxy spirals. 

From Fig.~3, a comparison of the H$\alpha$ map 
(Panel 1) and the radio map (contours in Panel 2) shows 
that the major axis of the ionised gas distribution is 
inclined to the axis of the radio jet by about $20^{\circ}$. At face value, 
this implies a misalignment between the radio jet ejection
axis and the normal to the plane of the putative central molecular torus,
which defines the orientation of the ionisation cones.
While this can be explained by models with accretion disk
precession or warps, we propose the following geometric model to account
for the misalignment (see Fig.~7). We begin by assuming that the inner jet is in fact 
aligned with the axis of an illumination cone. The cone axis is 
inclined from the normal to the circumnuclear gas disk 
by an angle $\phi$. In the assumption
that almost all the circumnuclear gas in confined to the disk and that the
disk scale-height is small (a thin disk), the emission line distribution
is determined by the ionised gas in the disk, which takes the 
form of a planar section through the illumination cone (a conic,
though only approximately, given the inhomogeneity and patchy
dust obscuration within the disk). The axis
of the EELR is then the projection of the illumination cone axis 
(i.e., the inner radio jet axis) onto the circumnuclear disk,
with the direction of projection along the normal to the disk plane.

We define two mutually rotated reference frames: a) a `disk frame'
with the circumnuclear disk in the XY plane and the normal to the
disk along the Z-axis, and b) a `sky frame', with the sky in the X'Y' plane
and the line of sight to the nucleus along the Z'-axis. 
We choose the X and X' axes to lie along the line of nodes 
of the disk, making it the common axis of rotation between the two frames.
We denote the rotation angle (anti-clockwise looking towards the origin) to
be $\alpha$. This angle is the inclination angle of the
disk . The jet axis is taken to be a radial vector with an angle 
from the Z-axis of $\phi$ ($\phi^{'}$) and from the line of nodes of  
$\theta$ ($\theta^{'}$) in the disk (sky) frames. Basic rotational 
transformations between spherical co-ordinate frames gives us 
relationships between these angles:

\begin{eqnarray}
\tan\theta^{'} & = & \frac{\cos\alpha\sin\theta\sin\phi - \sin\alpha\cos\phi}
{\cos\theta\sin\phi} \\
\cos\phi^{'} & = &\cos\alpha\cos\phi + \sin\alpha\sin\theta\sin\phi
\end{eqnarray}  
\vskip 0.2cm 
 
Since the axis of the EELR (i.e., the projection of the jet axis onto the
XY plane in the disk frame) is essentially aligned with the line of nodes, then
$\theta\approx0^{\circ}$ (see \S4.1). Taking $\alpha=42^{\circ}$ (the inclination of
the circumnuclear disk) and $\theta^{'}\approx-16\fdg5$ (the angle between
the radio jet axis and the line of nodes in the sky frame, which is
equal to the PA of the line of nodes, since the jet is aligned north-south),  
we can solve for $\phi$ using Eqns.~5 and 6:

\begin{equation}
\tan\phi = \frac{\sin\alpha}{\cos\alpha\sin\theta - \cos\theta\tan\theta^{'}}
=\frac{\sin\alpha}{\tan\theta^{'}}
\end{equation}

\noindent which gives $\phi\approx70^{\circ}$. In other words, the jet axis
is inclined to the plane of the disk by $90^{\circ}-\phi\approx 20^{\circ}$
Using Eqn.~6, we calculate the angle between the jet axis and 
line-of-sight $\phi^{'}\approx72^{\circ}$.

In the limit of a very thin disk and a well-defined illumination cone,
the opening angle of the cone can be estimated from the apparent opening
angle of the ionised section of the disk. Taking this to be between $70^{\circ}$ to
$90^{\circ}$, simple geometric arguments give cone opening angles of
around $80^{\circ}$ to $100^{\circ}$, typical of Seyfert ionisation cones. 

Finally, an estimate can be made of the thickness of the disk. If we assume
that the kinematically disturbed inner plume coincides with the projected 
length $l$ of the outflow as it interacts with the disk, then the disk 
thickness $D$ is given by:

\begin{equation}
D \: = \: \frac{2l}{\sin\phi^{'}}\:\sin\phi \: \approx \: 150 \textrm{ pc} 
\end{equation}
   
\noindent with $l=0\farcs5$ ($75$ pc) and the values of $\phi$ and 
$\phi^{'}$ estimated above. This number is quite 
plausible, though some caveats must be noted. Real disks are 
not sharply edged but likely to
have an exponential vertical gradient in gas density, while the gas 
distribution in the disk will be highly clumpy and non-uniform. 
Both these effects would lead us to underestimate the disk thickness. 
Another important consideration is that the outflow
could disrupt the orderly structure of the cold disk gas. 

\begin{figure*}
\label{geomodel}
\centering 
\includegraphics[width=0.9\textwidth]{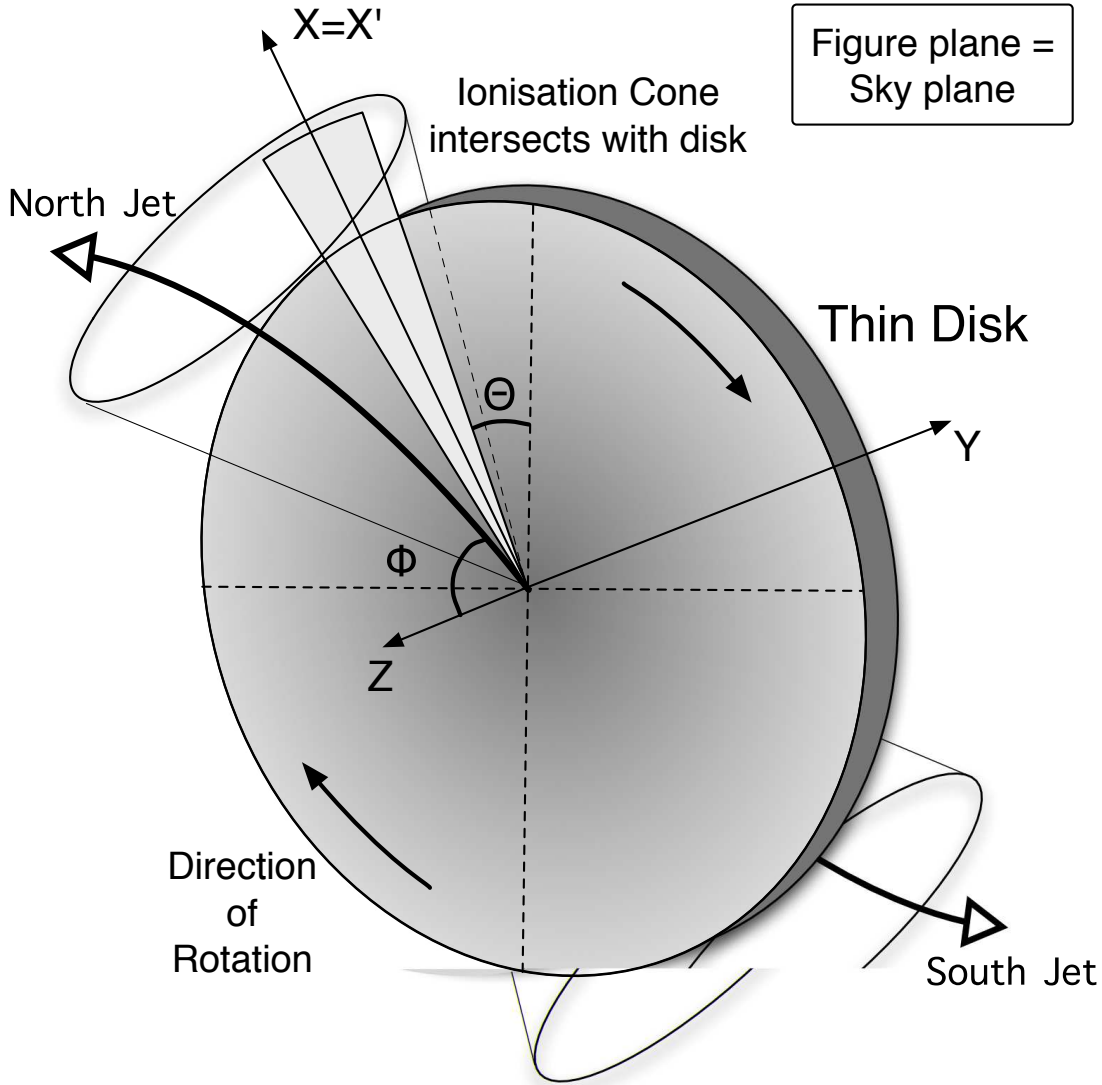}
\caption[NGC 2110: Geometric model]
{A schematic diagram of the circumnuclear region of NGC2110. A inclined rotating disk containing gas and dust
is intersected by a biconical radiation field. A radio jet emerges along the axis of the bi-cone but bends away from the 
disk. The various angles and axes are defined in \S6.}

\end{figure*}

The nature of the circumnuclear disk has also implications for the
radio source morphology as well. \citet{ulv83} consider possible causes 
for the symmetric bending of the jet in NGC 2110: 
a) ram pressure forces from the rotating ISM, 
b) static pressure gradients in the bulge and, c) precession of the
jet axis. It is possible to rule out ISM ram pressure as the main cause of
the jet bend, since the inferred direction of the disk rotation
is opposite to that required to bend the jet to its current shape, except
for the unlikely case where the tangent plane to the jet beam
sweeps across the LOS plane over the extent of the jet \citep{fiedler84}. 
While precession cannot be excluded, it seems unlikely given the constant jet
alignment down to parsec scales \citep{mundell00}. Given the strong gradient
in gas density perpendicular to the circumnuclear disk and the
fact that the jet bends towards the minor axis of the disk, we propose that
the jet bends as a result of refraction in the vertical pressure 
gradient of the disk.  

With this configuration for the jet and the disk, the inner emission line 
plume marks the location of higher density outflowing gas confined 
to the disk, ionised mostly by UV radiation from the AGN. 
We find little evidence for gas entrained into the jet itself: for
most of its extent, the jet and its terminal lobes are 
free of associated high velocity emission line material. 
This may be compared to the view of the jet interaction in Mkn 78 
\citep{whittle04}, where large amounts of line-emitting gas
are embedded and carried along by the jet flow. With this 
simple model, we now explore the general energetics of the NLR and 
radio jet from a dynamical standpoint.  

\section{Dynamical Study} \label{dyns} 

NGC 2110 hosts a significant radio jet and a relatively luminous
AGN, both of which can drive a fast moving outflow, either through
the action of relativistic pressure, ram pressure or the momentum
imparted by nuclear radiation, i.e, radiation pressure. In this section, we estimate the capacity
of each of these mechanisms to provide the total energy 
and momentum of the ionised outflow, in order to gauge their importance.

\subsection{Emission Line Region Properties}

If the gas is primarily photoionised (\S5), the masses, momenta and energies of 
the inner plume can be derived from basic nebular photoionisation theory
\citep{osterbrock}. The total flux in H$\beta$ measured from the STIS spectra and
corrected for an extinction of $A_{V} = 1.1$ magnitudes is
$F_{H\beta} \approx 2.4\times10^{-14}$ erg cm$^{-2}$ s$^{-1}$. From this,
the mass in emission line gas can be estimated using the following relation:

\begin{equation}
M_{em} \approx 171\, F_{H\beta}\,d^2 \, n_{em}^{-1}
\end{equation}

\noindent giving $M_{em} \sim 1.5\times10^{4}$
solar masses, where we have adopted a hydrogen number density of
$n_{em} = 1200$ cm$^{-3}$, derived from the ratio of the 
[S II]$\lambda\lambda6716,6731$ doublet \citep{ferruit99}.  Approximating the plume as a
cuboid of volume V$_{em} \approx 0.2\times0.2\times0.5$ arcsec$^3$ 
(or $6.76\times10^4$ pc$^3$), we can estimate the filling factor of ionised clouds:

\begin{equation}
f\!f_{em} \approx \frac{M_{em}}{n_{em}\; m_p \textrm{V}_{em}}
\end{equation}

\noindent to be $8\times10^{-3}$, a typical value for NLR clouds. Combining $ff_{em}$
with an estimate for the column density of the NLR clouds 
$N_{em} \sim 2\times 10^{21}$ cm$^{-2}$ from X-ray absorption models
\citep{ferruit99}, we derive typical covering factors of ionised gas in the NLR:

\begin{equation}
C_i \approx ff_{em}\, \frac{3}{2} \, \frac{n_{em}\, l}{N_{em}} \,\, \sim \;\; 1
\end{equation}

In addition, a simple estimate can be made of the age of the interaction 
from the approximate time taken by the emission line gas to cross the region of the plume.
With a characteristic velocity $v_{em} \approx 400$ km s$^{-1}$, we get a
timescale $t_{age} \approx l/v_{em} \sim 2\times10^{5}$ years.
This age estimate is essentially an upper limit if the gas in the plume
is decelerating with distance from the nucleus - a plausible interpretation
of the bulk velocity profile of the plume. 

Combining $M_{em}$ with the kinematics of the [O III] line and accounting
for a projection angle to the line of sight of $\phi=70^{\circ}$, 
the total (translational and internal) emission line
kinetic energy is $E_{ke} \sim 10^{52.3}$ erg and a bulk momentum is
$G_{em} \sim 10^{44.8}$ dyne-s. $E_{ke}$ may be compared to the 
equipartition energy in the radio source \citep[$\sim 10^{53}$ erg,][]{ulv83}
or the integrated energy from nuclear UV radiation over the
lifetime of the interaction ($\approx L_{ph} \times t_{age} = 5 \times 10^{56}$ erg).
Energetically, both the jet and nuclear radiation are capable of supplying the
energy of the outflow, with the nuclear radiation field dominating, by far, the
energy budget of the system. 

Assuming that the relativistic particles and magnetic fields of the
radio-emitting plasma are in equipartition, various physical parameters
of the radio jet can be estimated using relations from \citet{miley80}
and radio measurements from \citet{ulv83}.
Radio source pressures are modest, with
$\log P_{rel} \sim -9.0$ dyne cm$^{-2}$. These may be compared to the
pressures of the emission line gas and the extended hot gas in the NLR
responsible for the soft X-ray emission. 
Using $n_{em}$ and a temperature 
$T_{em} \approx 13600$ K from the [O III] $\lambda\lambda 4363/5007$ 
ratio \citep{ferruit99}, we determine the ionised gas pressure to be 
$\log P_{em} \sim -8.6$ dyne cm$^{-2}$. \citet{evans06} estimate
the X-ray emitting gas to be at a pressure of log $P_{X} \sim -8.5$
Both the emission line gas and hot gas appear to be slightly over-pressured with
respect to the radio jet plasma, but, given the uncertainties of a few
associated with such calculations, this is basically in agreement
with the conclusion of \citet{ulv83} that the phases are in 
rough pressure balance. Note that there is no good evidence 
that equipartition is a valid condition in such weak Seyfert jets.
Studies of FRI jets in radio galaxies suggest that they are particle-dominated, 
probably due to entrainment of a heavy thermal ions, and quite out of
equipartition \citep[see the review by][]{mcnamara07}. It is unclear whether
this scenario will also apply to small Seyfert jets in gas-rich circumnuclear environments.
However, given the similarity of the pressures across phases in NGC 2110, we
assume that equipartition is probably valid and proceed accordingly.
  
\subsection{Possible Acceleration Mechanisms}

Over the lifetime of the outflow, the forces that accelerate the outflow
have to be able to supply the bulk momentum measured in the inner plume.
The momentum supplied to the gas per unit time, the momentum flux
$\Pi_{em} \equiv G_{em}/t_{age}$, must be matched by any relevant accelerating
mechanism. From the estimates above, $\log \Pi_{em} \approx 32$ dyne.

\bigskip
\noindent A.) Relativistic Pressure

Is the radio source pressure capable of providing
the momentum of the emission line gas? We envision the jet
pressure acting on a sheath of ionised gas in contact with the
jet surface and accelerating the gas outward from the surface.
The area of this sheath $A_{beam}$ is approximately the surface area of a cylindrical
beam of length $l$ and cross-sectional radius of the jet $R_j \approx 70$ pc,
measured from the radio map. The relativistic pressure acting over this area gives a momentum
flux of $\Pi_{rel} \approx P_{rel} A_{beam}$, or $\log \Pi_{rel} \sim 31.8$
dyne. Within the considerable uncertainties, $\Pi_{rel} \sim \Pi_{em}$ and the relativistic pressure
could, in principle, accelerate the emission line gas.

\bigskip
\noindent B.) Jet Ram Pressure

The ram pressure of the fast moving jet flow as it propagates from the nucleus
will accelerate clouds that lie along its path. While the physics of interactions
between a jet and ISM clouds involves complex hydrodynamic processes such as 
shocks, cloud-breakup and ablation, we take a broader view and ask simply whether
a jet with a ram pressure $P_{ram} \equiv \rho_j V_j^2$ can satisfy the constraints
of $\Pi_{em}$, where $\rho_j$ is the jet density and $V_j$ is the jet velocity.

A constraint on these jet parameters comes from the morphology of the jet.
We have argued that the most likely explanation for the
symmetrical bend in the jet is the action of a disk pressure gradient.
Modelling the disk as a slab with a scale height much smaller than the
disk plane scale length, we can apply the following condition for significant 
jet deflection from \citet{hvb81}:

\begin{equation}
\rho_j V_j^2 \approx \frac{\gamma}{\gamma -1 } P_s
\end{equation}

\noindent where $P_s$ is the disk pressure at the sonic point of the jet
and $\gamma$ is the polytropic equation of state of the jet material ($=4/3$ for
a relativistic gas). Assuming that the line emitting gas is in rough pressure 
equilibrium with the hot atmosphere of the galaxy, i.e, $P_{em} \sim P_s$,
we get jet ram pressures of log $P_{ram} \sim -8.0$ dyne cm$^{-2}$.
This pressure acting over the area of cross-section of the jet gives a 
force of $\Pi_{ram} \approx \pi R_j^2 P_{ram}$ or log $\Pi_{ram} \sim  33$ dyne.
Interestingly, requiring the jet to bend in the disk pressure gradient also implies 
a jet velocity and density that can easily accelerate the outflow by ram pressure. 

Taking a fiducial proton number density for the jet of $\rho_j/m_p = 10^{-2}$
cm$^{-3}$, we get $V_j \sim 10^{-2}c$, which is highly sub-relativistic.
In other words, the requirement of  substantial buoyant force on the jet material 
in the galaxy's atmosphere implies  that the jet is either highly underdense or highly 
subrelativistic.

\bigskip
\noindent C.) Radiation Pressure

The emission line gas in the plume is also illuminated by
a powerful ionising radiation field from the nucleus. The pressure from
this radiation is potentially a major source of momentum to the
ionised gas. We estimate its momentum flux from the photon luminosity of the nucleus 
and the covering factor of the inner plume $\Omega C_i \approx 0.02$:

\begin{equation}
\Pi_{rad} = \frac{\Omega C_i \, L_{ph}}{c}
\end{equation}

\noindent which yields $\log \Pi_{rad} \approx 31.5$, a little lower than $\Pi_{rel}$.
Within our uncertainties, radiation pressure is also capable of driving the 
nuclear outflow. However, the ability of radiation pressure to impart 
momentum to the ionised gas depends on nuclear distance simply because at smaller 
distances the covering factor  $\Omega C_i $ is greater.

\subsection{The Mass Outflow Rate}

The mass outflow rate in the jet is given by:

\begin{equation}
\dot{M}_{jet} \; = \;  \pi \rho_j V_j R_j^{2} \;  = \;  \pi P_{ram} R_j^{2} / V_j \; \approx \; 7.8\times10^{-4}/V_j  \quad \textrm{M}_{\odot} \textrm{yr}^{-1}
\end{equation}

{\noindent where we have adopted the value of $P_{ram}$ from the jet bend analysis above}.

The Eddington accretion rate for the SMBH in NGC 2110 is $\sim 5.3 \textrm{ M}_{\odot} \textrm{yr}^{-1}$, 
assuming a radiative efficiency of 0.1, while the mass accretion rate for the estimated bolometric 
luminosity of the AGN ($L_{ph}$) is $\dot{M}_{acc} \sim 8.8\times10^{-3} \textrm{ M}_{\odot} \textrm{yr}^{-1}$. 
Interestingly, the mass accretion rate of the SMBH is close to (possibly even lower than) the mass
outflow rate in the jet if we adopt our fiducial value for $V_j$. 
As long as the constraint on $P_{ram}$ is valid, this may suggest that the jet carries
away a substantial fraction of the mass accreted onto the SMBH. However, it is also possible that the
current AGN luminosity may have varied substantially over $t_{age}$, since the time of the ejection of the jet,
that the jet is highly underdense or modestly relativistic, or that the jet ram pressure estimate is too high. 
The large number of contributing factors makes the determination of the most likely cause for this
discrepancy difficult to ascertain.

\subsection{The Role of Gravity}

If any of the acceleration mechanisms discussed above are to accelerate gas in an outflow, 
it must overcome the local gravitational force of the galaxy's spheroid acting on the gas. 
The relative strengths of these two opposing forces can be estimated as follows. Consider
a spherical optically thick cloud with radius $R_c$, density $n_e$, located a distance
$d$ from the nucleus. The gravitational force on the cloud is:

\begin{equation}
F_{grav} \approx \frac{4 G m_{p} R_{c}^{3} n_{e} M(<d)}{d^{2}} \approx \frac{4 m_{p} R_{c}^{3} n_{e} V_{c}^2}{d}
\end{equation}

\noindent where $m_p$ is the proton mass, $M(<d)$ is the spheroid mass within $d$ and $V_{c}$ is the circular
velocity corresponding to $M(<d)$.  

The radiation force acting on this cloud is:

\begin{equation}
F_{rad} =  \frac{1}{4c} \frac{L_{ph} R_{c}^{2}}{d^{2}}
\end{equation}

\noindent where $L_{ph}$ is the nuclear luminosity and assuming that all nuclear radiation illuminating
the cloud is absorbed. The ram pressure force is:

\begin{equation}
F_{ram} =  \pi \rho_j V_j^2 R_c^2
\end{equation}

\noindent where we used the definition of $P_{ram}$ from above. 

The ratio of gravitational force to radiation force on the cloud is:

\begin{equation}
\frac{F_{grav}}{F_{rad}} \approx 8 m_p c \; \frac{V_c^2 N_{e} d}{L_{ph}} 
\end{equation}

\noindent where we have used the cloud column density $N_e \approx 2 n_e R_c$.  
Significant acceleration by radiation pressure can only occur if this ratio is $< 1$. Taking
estimates of $L_{ph}$ and $N_e$ from above, and assuming a constant $V_{c} \sim 300$ \kms\ 
(based on the H$\alpha$ rotation curve from \citet{ferruit04} as well as our own measurements of the southern
NLR's [O III] velocities), this criterion implies a critical radius
of $d\sim20$ pc beyond which radiation pressure cannot effectively accelerate the gas. The inner
plume deprojected size of $l \sim 150$ pc is much larger than this critical radius. This suggests that
radiation pressure is not a significant acceleration mechanism for most of the resolved inner NLR.

The ratio of gravitational force to ram pressure force is:

\begin{equation}
\frac{F_{grav}}{F_{ram}} \approx \frac{2}{3} \left ( \frac{V_c}{V_j}\right )^2 \frac{N_{e}m_p}{\rho_j d} 
\end{equation}

\noindent This ratio is inversely proportional to nuclear distance. If $P_{ram}$ is roughly
constant over the length of the jet, the ratio drops linearly with $d$, such that ram pressure 
begins to dominate over gravitational forces \emph{above} a critical radius of about 20 pc. 
Unlike radiation pressure, ram pressure of the jet flow may be an important acceleration 
mechanism in the plume. 

A treatment of relativistic pressure in the same fashion is difficult, since it depends critically on physics of
the boundary between the relativistic plasma and the ionised gas. Undoubtedly, gravity will have
an effect on the displacement of any gaseous component in contact with relativistic plasma, but
we forgo a discussion of relativistic pressure effects, since the jet ram pressure is likely to dominate
the acceleration of any gas that comes into contact with the jet flow.

\subsection{Gas Motions in the EELR}

\citet{ferruit04} argue that the asymmetry between the velocity profiles of the northern and southern EELR
can be explained if the northern part is disturbed by the jet, while the south is in normal rotation in a thin
disc. The low S/N of the line emission in our spectra on the scales of the EELR (nuclear radii $ > 1"$)
does not allow a full exploration of this phenomenon. We can, however, confirm that an asymmetry does
exist between the absolute peak velocity of the narrow emission in the NW and SE EELR along Slit A,
measured at and beyond $1"$ (see Fig.~4). We may speculate on possible causes for the asymmetry. 
In our model for the NLR, the jet escapes the disk where most of the observed emission line gas is confined
and therefore, direct interaction with the jet is unlikely to be the main cause of the asymmetry. This is supported
by the fact that the jet is quite symmetric in its structure, which belies the basic symmetry of the gas density
gradients that bend the jet. In this light, we propose two alternative hypotheses. 

A.) We have shown that radiation pressure probably cannot drive a strong outflow beyond few tens of parsecs.
However, it may still be able to influence the gravitational motion of AGN ionised gas on scales of the EELR. 
There is a strong difference in the mean ionisation between the northern and southern EELR \citep{wilson85a, ferruit04},
which implies a difference in the ionising field strength and possibly the contribution of radiation pressure. 
A detailed study of the conditions of the emission line gas and any ionisation-kinematic trends in the EELR
may test this idea, but this is beyond the scope of this work.

B.) The northern EELR has a pronounced dusty spiral arm that has no obvious counterpart in the south.
Such structures have been proposed as paths along which gas streams into the nucleus from the outer
parts of the galaxy \citep{maciejewski04, fathi06, davies09} due to spiral shocks. A comparison of the spiral
pattern with the kinematic residuals left after accounting for normal disc rotation, 
such as that presented by \citet{gonz02}, may provide evidence for this model.



\section{Summary} \label{discussion}

The circumnuclear environment of NGC 2110 consists of several components which each play a role in
shaping the structure and dynamics of the region and the NLR. We explore the outflow using 
well-resolved kinematics from STIS spectroscopy and constrain the influence of the circumnuclear
components using high resolution optical and radio imaging. Summarising our results:

\begin{enumerate}

\item A circumnuclear dusty disk plays a major part in shaping the appearance of the NLR. We
propose a model that explains the angular offset between the axis of the inner radio jet and the EELR
in terms of the biconical illumination of this dusty gaseous disk. We also argue that the strong,
symmetric bending of the jet is a consequence of the anisotropic mass and pressure distribution from the
disk. 

\item A high surface brightness emission line structure, which we call the ``inner plume'', is visible within the 
inner $\sim 100$ pc of the nucleus. The kinematics  of the inner plume reveal it to be
accelerated, broadened and disturbed -- the site of a localised nuclear outflow. 
Its geometry indicates that it is not aligned along the radio jet, but possibly skirts the main jet flow. 
An ionisation analysis indicates that the inner plume is photo-ionised by the central active nucleus.
Accounting for dust extinction in the disk, we conclude that the plume is possibly part of a extended
symmetric outflow on either side of the nucleus.

\item We explore the ionisation and kinematics of a bright compact nuclear knot, 
which has a size smaller than 20 pc and is barely resolved, even with HST. The low ionisation state of the knot
and its high velocities, which are continuous with those of the inner plume, reveal that the nuclear
outflow starts out on these small scales and then expands outwards. 
We speculate whether some of the kinematic structure in the knot may come from motion 
around the central black hole. 

\item Finally, we explore the dynamics of the outflow. Linking the jet bend to the pressure of the bulge atmosphere
leads to constraints on the jet density and velocity. We suggest that this jet is significantly
sub-relativistic and has sufficient ram pressure to drive the emission line outflow. Radiation pressure
from the active nucleus or radio source pressure can both also account for the 
energetics of the inner plume, but the increasing influence of the gravitational potential 
of the galaxy with nuclear distance restricts the importance of radiation pressure in accelerating
gas in the resolved inner plume. 

\end{enumerate}

\section{Acknowledgements}
DR acknowledges the support of the National Science Foundation through grants AST-0507483  and
AST-0808133. We thank Neil Nagar for providing updated radio maps for this work.
Based on observations made with the NASA/ESA Hubble Space Telescope, obtained 
from the data archive at the Space Telescope Institute. STScI is operated by the association of 
Universities for Research in Astronomy, Inc. under the NASA contract  NAS 5-26555.
The National Radio Astronomy Observatory is a facility of the National Science Foundation 
operated under co-operative agreement by Associated Universities, Inc.



\end{document}